# Fairness Perceptions of Algorithmic Decision-Making:

# A Systematic Review of the Empirical Literature

Christopher Starke[1], Janine Baleis[1], Birte Keller[1] & Frank Marcinkowski[1]

[1] Heinrich Heine University Düsseldorf

**Abstract**

Algorithmic decision-making (ADM) increasingly shapes people's daily lives. Given that such autonomous systems can cause severe harm to individuals and social groups, fairness concerns have arisen. A human-centric approach demanded by scholars and policymakers requires taking people's fairness perceptions into account when designing and implementing ADM. We provide a comprehensive, systematic literature review synthesizing the existing empirical insights on perceptions of algorithmic fairness from 39 empirical studies spanning multiple domains and scientific disciplines. Through thorough coding, we systemize the current empirical literature along four dimensions: (1) algorithmic predictors, (2) human predictors, (3) comparative effects (human decision-making vs. algorithmic decision-making), and (4) consequences of ADM. While we identify much heterogeneity around the theoretical concepts and empirical measurements of algorithmic fairness, the insights come almost exclusively from Western-democratic contexts. By advocating for more interdisciplinary research adopting a society-in-the-loop framework, we hope our work will contribute to fairer and more responsible ADM.

*Keywords*: Algorithmic Decision-Making, Fairness Perceptions, Justice, Discrimination, Literature Review

Correspondence to: Christopher Starke, christopher.starke@uni-duesseldorf.de

# 1 Introduction

Algorithms increasingly shape people's daily lives by making important decisions in public administration (Wirtz et al., 2019), in lending (Lessmann et al., 2015), in the legal system (Završnik, 2020), and in medicine (Rajkomar et al., 2019). While algorithmic decision-making (ADM) can lead to faster and better decision outcomes (Lepri et al., 2018), it often includes a downside: due to biased input data or faulty algorithms, unfair ADM systems have been proven to systematically reinforce racial or gender stereotypes, marginalize minorities, or flat-out denigrate certain members of society (Veale & Binns, 2017; Žliobaitė, 2017). The famous example of the COMPAS algorithm, which disproportionately assigned a higher risk score of recidivism to black defendants than to white defendants, is evidence of the existence of algorithmic discrimination (Chouldechova, 2017).

Fairness has become a key element in developing sociotechnical algorithmic systems to counter such detrimental results (Hutchinson & Mitchell, 2019). Algorithmic fairness is endorsed as one of the four main principles for trustworthy Artificial Intelligence (AI) by policy institutions like the OECD (2019) and the EU (Artificial Intelligence High-Level Expert Group, 2019), and it has been featured in more than 80 percent of guidelines for AI ethics (Jobin et al., 2019). However, addressing the societal implications of (un)fair ADM requires more than mere technological solutions (Barabas et al., 2020; Sloane & Moss, 2019): designing and implementing human-centric algorithms calls for a thorough empirical understanding of when and why citizens *perceive* ADM to be (un)fair. Here, we present the first authoritative systematic literature review mapping the existing empirical insights on *fairness perceptions of algorithmic decision-making*.

This paper synthesizes the interdisciplinary results of 39 empirical studies, incorporating over 25,000 unique observations of citizens' fairness perceptions of ADM. We apply a state-of-the-art approach to capturing both academic and gray literature by combining a systematic online search with a "pearl growing" strategy of sequential citation searches. Screening more than 2,500 entries, two coders systematically identified the relevant literature in two subsequent coding steps. We systemize the existing literature along four main dimensions of perceived algorithmic fairness: (1) algorithmic predictors, (2) human predictors, (3) comparative effects (human decision-making (HDM) vs. algorithmic decision-making), and (4) consequences of ADM.

## 2  Bias in Algorithmic Decision-Making

Algorithms have the potential to reduce human biases in decision-making processes because they do not grow tired or inattentive, they have no agency, and they are not distorted by emotional factors (Lee, 2018). For instance, in South Korea, algorithms were used to relocate ambulance units so that more people could receive help within five minutes of making an emergency call (Nam, 2020). Furthermore, in their seminal study, Bansak et al. (2018) used a machine learning (ML) model to assign refugees to resettlement locations, leading to improved integration outcomes. However, in many cases, AI-based systems decreased fairness (Barocas & Selbst, 2016). For instance, a recent report by the research institute AI NOW revealed that ADM systems have arbitrarily excluded citizens from food support programs, mistakenly reduced their disability benefits, or falsely accused them of fraud (Richardson et al., 2019).

Biases in ADM are often unintended and can be explained by different reasons. They can occur in collecting and processing input data but also in selecting and specifying an ML system (Veale & Binns, 2017). First, in terms of training data, ADM systems that learn from historical input data are likely to reproduce or even exacerbate existing biases in society, often with harmful outcomes for minority groups (Köchling & Wehner, 2020; Lepri et al., 2018). Furthermore, data about individual or group features may be incomplete or unreliably measured, leading to underrepresentation or overrepresentation of certain groups (Köchling & Wehner, 2020). Second, algorithms may discriminate if they are carelessly selected, designed, and specified, as some ML models may perform fairly on some specific tasks but unfairly on others. For instance, random forests or neural networks are better suited to capture the synergy among variables than are linear regression models (Veale & Binns, 2017).

However, reducing such bias is not merely a technical challenge (Lepri et al., 2018; Wong, 2020); problematic impacts are not necessarily a result of biased input data and faulty algorithms, but can also occur in the implementation of an ADM system (Köbis et al., 2021)—e.g., when the system violates privacy rights or is used for sensitive decisions that should not be made by an algorithm in the first place. Wong (2020) further pointed out the importance of appropriate fairness concepts, arguing that "one can dispute whether an algorithm is fair by questioning the idea of fairness underlying the 'fair' algorithm in question" (Wong, 2020, p. 227). To better understand this contentious concept, the following section focuses on the different existing notions of algorithmic fairness.

## 3    Concepts of Algorithmic Fairness

In recent years, the concept of fairness has regained prominence as a core objective in designing AI (Jobin et al., 2019). The term 'algorithmic fairness' generally means that decisions made by an algorithm should not produce unjust, discriminatory, or disparate consequences (Shin & Park, 2019). Two broad approaches to algorithmic fairness can be identified: first, literature that formalizes algorithmic fairness and derives mathematical definitions (Gajane & Pechenizkiy, 2017; Verma & Rubin, 2018; Žliobaitė, 2017); second, literature that draws on philosophical and social-science concepts of human fairness and applies these concepts to algorithms (Binns, 2018a, 2018b; Marcinkowski & Starke, 2019).

Different formal definitions of algorithmic fairness compete in the former strand of literature. Reviewing more than 20 different definitions, Verma and Rubin (2018) found that formal concepts can largely be clustered into three categories (for an elaborate historical discussion of fairness definitions, see Hutchinson & Mitchell, 2019). First, *statistical measures* (e.g., statistical parity, Dwork et al., 2012; equalized odds, Hardt et al., 2012) are based on different calibrations of predicted probabilities, predicted outcomes, and actual outcomes. Second, s*imilarity-based measures* (e.g., fairness through awareness, Dwork et al., 2012) are based on the premise that similar individuals should be treated similarly, regardless of their classification in various specific groups. Third, *causal reasoning* (e.g., counterfactual fairness, Kusner et al., 2017) assumes that structural equations can be used to estimate the effects of sensitive attributes and to design algorithms that ensure tolerable discrimination levels due to these attributes. Adding to the list of formal definitions, Zafar et al. (2017) introduce preference-based fairness (e.g., preferred treatment, preferred impact), conceived as a predictor that increases benefits for a group compared to another predictor.

The plethora of existing formal fairness definitions indicates that they refer to different notions of fairness, but these conceptions are also often incompatible with each other (Kleinberg et al., 2017). Thus, as different fairness trade-offs emerge, several authors have highlighted the importance of the social context when assessing appropriate understandings of algorithmic fairness (Lepri et al., 2018; Wong, 2020).

The second strand of literature addresses the fairness of ADM from a different theoretical perspective. In two seminal papers, Binns (2018a, 2018b) drew on moral and political philosophy to outline a concept of algorithmic fairness. He discussed how egalitarianism—the belief in equal treatment of people and the equal distribution of fundamental rights and goods—can inform theoretical notions of algorithmic fairness. For instance, Binns (2018a) pointed out that algorithmic (un)fairness cannot only be assessed on the grounds of unequal distribution; instead, it should also consider how inequality is produced. Thus, an algorithm may need to consider additional information to make fair decisions, (e.g., about historical structural injustices against certain minorities in society).

Other authors have come to similar conclusions despite using a different theoretical approach (Grgić-Hlača, Zafar, et al., 2018; Marcinkowski & Starke, 2019). Drawing on the seminal distinction in organizational justice literature (Greenberg, 1987, 1990), they have assessed algorithmic fairness according to four different dimensions. First, *distributive fairness* refers to the non-discriminatory allocation of resources based on equality, equity, or need (for a more detailed discussion, see Deutsch, 1975). Second, *procedural fairness* indicates that decision-making is based on fair criteria, such as revocability or consistency (Leventhal, 1980). Third, *informational fairness* involves the transparency of ADM systems—i.e., whether or not it provides explanations for decision outcomes (Greenberg, 1993). Fourth, *interpersonal fairness* is achieved if an ADM refrains from using protected data and respects privacy rights (Greenberg, 1993).

In their interdisciplinary approach, Gajane and Pechenizkiy (2017) made an attempt to contrast the formal fairness definitions dominating the computer science literature with the social science fairness theories. They pointed out that the social science concepts of *equality of*

*resources* and *equality of capability of functioning* have not been adequately addressed in the ML literature. Equality of resources describes a mechanism to ensure that people have a fair, if not equal, distribution of resources that allows them to individually choose those goods that they want (Dworkin, 1981). Equality of capability of functioning calls for equalized opportunities with no regard to unequal social or natural endowments and compensation for those unequal powers (Sen, 1990).

This reasoning adds to several authors' argument that fair predictions by algorithms cannot be made without considering social questions. ADM systems do not operate in a vacuum but rather need to be calibrated to the specific social context. As Shin and Park (2019, p. 279) stated: "What establishes an algorithm system as a socio-technical system is that it is generated by or related to a system adopted and used by social users in societies." As it is ultimately humans who are affected by the decisions made by ADM, several authors advocate for a more human-centric approach to researching algorithmic fairness in order to ensure that ADM systems are legitimized (Grgić-Hlača, Redmiles, et al., 2018; Marcinkowski et al., 2020; Srivastava et al., 2019). Many empirical studies have addressed this call by empirically investigating human perceptions of algorithmic fairness. It is that growing literature that we review in this paper.

## 4 Method

We conduct a systematic review of the empirical literature that investigates people's perceptions of algorithmic fairness. We adopt the approach recommended by Petticrew and Roberts (2006) that was tailored to social science literature. It outlines seven steps to ensure a thorough literature review based on predefined and transparent criteria. The first step develops the *research question* or *hypothesis* to be answered by the review. Second, *inclusion criteria* for studies are defined. Third, a *comprehensive literature search* is conducted. Fourth, a *screening of the results* is required to classify the literature according to its relevance. Fifth, a *critical evaluation* of the selected studies is carried out before, sixth, the *contents are summarized* and, seventh, the *results of the review are disseminated*.

## 4.1  Establishing the research question

A precise research question is vital for a systematic literature review (Booth et al., 2016). We apply the 'PICOC' method (Petticrew & Roberts, 2006, p. 44) to break down the research question into the following five individual components: **P**opulation, **I**ntervention, **C**omparison, **O**utcome, and **C**ontext. This approach has been proven to help identify potential search terms for subsequent searches in databases (Booth et al., 2016). We define each component in relation to the central aim of this literature review. First, the *population*—the subjects of the question—includes all individuals, irrespective of sociodemographic characteristics. Second, the *intervention* is defined as fairness in and through algorithms. Third, we did not specify a *comparison*, as it was not considered useful to include additional interventions. Fourth, the *outcome* involves individual fairness perceptions about ADM. Fifth, the *context*—conceived here as a country-specific and domain-specific setting—was not narrowed down, thereby making the literature review global in scope. Using these defined criteria, we derive the following research question: *How do individuals perceive the fairness of algorithmic decision-making?*

## 4.2  Inclusion Criteria

The keywords in the research question ('*fairness*', '*algorithmic*') provide the basis for the search terms. For additional terms for the subsequent search, we adopted the 'pearl-growing' method (Booth et al., 2016). It draws on relevant articles ('pearls') to identify further relevant search terms or keywords. Therefore, 'pearls' need to be closely related to the main research interest of the literature review. For this study, we initially identified three articles as 'pearls' because they are widely cited in the literature: Binns et al. (2018), Grgić-Hlača, Redmiles et al. (2018), and Lee (2018).

As all these articles correspond with our research interest, we added relevant keywords to the search terms we derived from our research question (see Table 1).

**Table 1.** Selected keywords.

| Keywords through **PICOC** | Keywords used in 'pearls' |
|---|---|
| - fairness | - justice |
| - perception | - discrimination |
| - algorithmic decision-making | - machine learning |

Subsequently, we clustered the search terms into two components: terms referring to ADM, and terms referring to the theoretical concept of fairness (see Table 2).

**Table 2.** Search terms for literature searches.

| Component 1 | Component 2 |
|---|---|
| - big data | - fair* |
| - artificial intelligence | - unfair* |
| - machine learning | - just* |
| - algorithm* | - discrimina* |
|  | - bias |
|  | - disparate |

We inputted these search terms into the following Boolean operators (Lefebvre et al., 2008): ("big data" *OR* "artificial intelligence" *OR* "machine learning" *OR* "algorithm*") AND ("fair*" *OR* "unfair*" *OR* "just*" *OR* "discrimina*" *OR* "bias" *OR* "disparate").

If the predefined combination of search strings appeared in the title of a publication, we included that publication in the preliminary sample, which we then used for the first screening process. Due to the recent surge of literature on algorithmic fairness, we adopted the reasoning suggested by Favaretto, De Clercq, and Elger (2019) and only included studies written in English and published since January 2010.

### 4.3 *Comprehensive Literature Search*

We selected the final sample of empirical studies through a stepwise process (see PRISMA chart, Figure 1). As fairness of ADM has been investigated in different research disciplines, the selection of the databases was based on thematic classification (see Table 4 in the supplementary material). We applied the Boolean logic to the following electronic databases: Web of Science, PsycINFO, IEEE Xplore, and Scopus. Although most systematic reviews only include articles from peer-reviewed procedures and leave out other publications (Jungnickel, 2018), our study also includes so-called 'gray literature'—that is, working papers, pre-prints, or reports—as they account for recent research efforts. We used Google Scholar to manually search for relevant publications by searching for publications by key institutions (e.g., AI NOW, AlgorithmWatch).

Applying the search terms to the selected databases, we identified a total of 4,045 contributions (15 March 2021), of which 2,467 remained after filtering for duplicates. In a subsequent research step, we identified 99 potentially relevant articles through the manual search of gray literature. For all publications, we extracted the title, authors, keywords, journal information, and abstract.

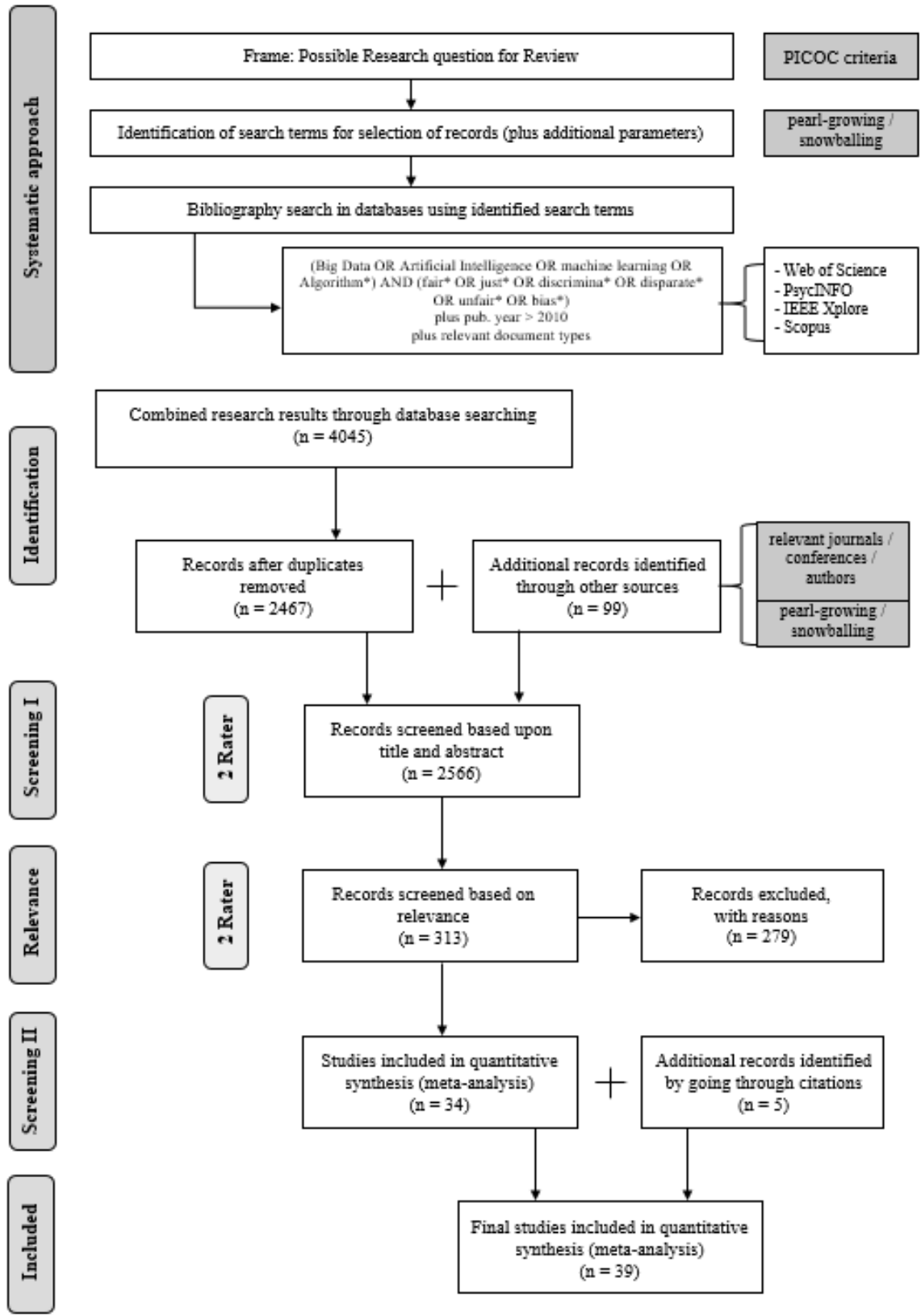

**Fig. 1.** Flowchart documenting the selection process adopted by Moher et al. (2009)

*4.4    Screening & Critical Evaluation*

For the initial screening, two expert raters examined the 2,566 publications based on their titles and abstracts. The screening was twofold. In the first step, for a publication to be considered for further analysis, both raters had to agree that it generally addressed the research interest. In this step, 313 publications were selected after initial screening of the title and abstract. In the second step, all publications were rated in terms of the following two criteria: applicability to the research question of the review, and relevance of the empirical method. For each measure, the raters could assign 1 = *applies* or 0 = *does not apply*. All publications that were rated 1 on both criteria were selected for the literature review. This filtering process resulted in a sample consisting of 34 publications. In a final step, we used Google Scholar to scan all publications that cited at least one of the selected 34 articles, identifying five other relevant publications that were included in the final sample.

We tested the reliability of the selection process with a sample of 72 abstracts that were assessed by both raters. Based on Higgins and Deeks's (2008) recommendation, we calculated the consistency measure Cohen's Kappa ($\kappa$) to ensure coherence of the coding results between the two raters. The first coding process resulted in a Cohen's $\kappa$ value of .55, which reflects "fair agreement" (Higgins & Deeks, 2008, p. 155). To improve the reliability, both raters discussed the examples on which they diverged. In the next step, a second test with 39 abstracts was conducted, leading to a Cohen's $\kappa$ score of .74, which is considered "good agreement" (Higgins & Deeks, 2008, p. 155).

# 5 Results

To start, we outline the descriptive results of the 39 studies included in the literature review. Then, we shed light on the underlying theoretical concepts of fairness and the specific measurements used in the empirical studies. We then reveal the main insights by clustering the empirical results from the existing literature on individuals' perceptions of algorithmic fairness into four main categories: *algorithmic predictors*, *human predictors*, *comparative effects* (HDM vs. ADM), and *consequences of ADM*.

## 5.1 *Descriptive Results*

The descriptive results (shown in Table 3 in the supplementary material) indicate high homogeneity in terms of the national context. Data on citizens' perceptions of algorithms' fairness was almost exclusively collected in Western democracies: 23 studies were conducted in the United States (US), two in the United Kingdom (UK), two in the Netherlands, one in Germany, one in South Korea, and one in China, with four studies collecting data in multiple countries[1].

Looking more closely at the empirical methods used to investigate citizens' perceptions of algorithmic fairness, we find great diversity. Seven studies used a qualitative design, 22 studies used quantitative methods (either surveys or experimental designs), and ten studies combined qualitative and quantitative approaches in mixed-method designs.

With regard to the different domains within which the studies were located, the descriptive results reveal a focus on the criminal justice system as an area of application (eleven

---

[1] Five studies do not indicate the country in which the data was collected.

studies)—most prominently pretrial risk assessment, such as the famous example of COMPAS. Moreover, emphasis is also given to work-related decisions (seven studies), especially in hiring. Other domains include news recommendations, allocation of donations, university admissions, loan decisions, and targeted advertisements.

**5.2** *Concepts of (Algorithmic) Fairness*

Many studies drew from theoretical fairness concepts developed for human interactions and adapted them to the context of ADM systems. Thus, we find great heterogeneity in relation to algorithmic fairness notions (see Table 3 in the supplementary material). The existing literature primarily focuses on the perceived fairness of decision outcomes—i.e., distributive fairness, although many computer science studies do not use this term explicitly. Several of those studies distinguished between different formal definitions of algorithmic fairness (e.g., demographic parity, equalized odds) (Verma & Rubin, 2018) or among equality, equity, and need (Deutsch, 1975). Thus, they conceived of algorithmic fairness in distribution norms. Going beyond distributive fairness, several studies investigated the fairness of algorithmic processes—i.e., procedural fairness.

However, conceptual ambiguity exists. Some studies from the computer science literature defined procedural fairness more narrowly as the inclusion of selected (sensitive) input features. However, procedural fairness was typically conceptualized in broader terms in social science studies, also addressing such criteria as the consistency or the revocability of decisions (Leventhal, 1980). Drawing on organizational justice literature (Greenberg, 1990), some studies also included interactional fairness and informational fairness, both involving the social side of fairness. Another aspect that received some attention in qualitative studies is equality of opportunity, accounting for inequalities in access to resources.

In sum, the results reveal a varied research field with regard to the concepts of algorithmic fairness. Most approaches seem to be based on ad-hoc applications of existing fairness concepts that were initially developed to assess fairness in human interactions. The empirical literature lacks a coherent, multi-dimensional theoretical concept of perceived algorithmic fairness.

### 5.3 *Measurements of Algorithmic Fairness*

Most quantitative studies used a two-step process to measure fairness perceptions of ADM: first, respondents were confronted with an ADM process or an ADM outcome, and then they were asked about their perceived fairness. However, the results show much diversity in the measurement of algorithmic fairness (see Table 3 in the supplementary material). Seven studies simply used single items measured on 5-point or 7-point Likert scales, along the lines of "how fair did you perceive the decision?". A total of 16 studies drew from existing fairness scales (e.g., Colquitt & Rodell, 2015) and adapted them for algorithmic decision-making. Three studies indirectly gauged fairness via stated preferences for one ADM over another. In contrast, Pierson (2017) predefined a fair distribution of resources and asked participants' approval on a 7-point Likert scale.

In sum, all studies in this literature review either used simple single-item measures of perceived fairness or adopted fairness scales that were initially designed for human decision-making. A validated multi-dimensional scale measuring fairness perceptions of ADM is lacking.

**5.4**  *Algorithmic predictors of perceived algorithmic fairness*

A significant strand of literature investigates how the technical design of an ADM system affects people's fairness perceptions. Several qualitative studies tapped into people's general notions of algorithmic fairness. Binns et al. (2018, p. 9) found somewhat mixed results: some respondents perceived "the *very idea* of an algorithmic system making an important decision on the basis of past data [...] unfair", while other participants argued that algorithms are by definition impartial and can therefore only vary in their degree of accuracy. However, the results obtained by Shin and Park (2019) suggested that fairness is a crucial factor when evaluating algorithms and that a respondent's understanding of algorithmic fairness primarily involves notions of indiscrimination, impartiality, and accuracy. Along similar lines, Dodge et al. (2019) found that respondents evaluated the fairness of an algorithm based on: the features that are included in the model, absent features, errors in the algorithm, and flawed or even false input data.

With its focus on recommender systems, the study by Smith et al. (2020) revealed that algorithmic fairness is perceived as more problematic in some domains than in others. For instance, discrimination by ADM in areas such as housing or job recommendations is viewed as more harmful than in areas such as music or movie recommendations (e.g., Spotify, Netflix). The results put forth by Koene et al. (2017) indicated that fairness is perceived to be highly context-dependent, with the authors concluding that there is no "(...) unique, globally approved, definition of fairness" (Koene et al., 2017, p. 2).

Going beyond investigating people's basic understanding of algorithmic fairness, a large group of studies investigates how people's perceptions of fairness are related to the plethora of existing formal definitions of fairness—in other words, the outcome distributions yielded by algorithms that people evaluate to be the fairest. This is particularly intriguing, as some fairness definitions cannot coexist (Kleinberg et al., 2017). In three studies, Lee and colleagues tested people's fairness perceptions regarding the allocation of resources based on equality, equity, or efficiency. Using qualitative interviews, Lee and colleagues looked at algorithms that match food donations with non-profit organizations (M. K. Lee et al., 2017; M. K. Lee, Kusbit, et al., 2019). They found much variation in the preferences for the three fairness concepts (equality, equity, efficiency), both within and across different social groups differently impacted by the decision. In a qualitative lab study, Lee and Baykal (2017) investigated the division of rents, household chores, snacks, or credit for a game outcome. Most respondents considered an outcome as fair when it mirrored their input (equity), yet some respondents also believed that an equal allocation of tasks or resources was fair, emphasizing moral norms such as self-sacrifice (equality).

While these qualitative studies focused on very basic and fundamental fairness concepts, a stream of quantitative studies tested more nuanced notions of fairness. In an experiment using criminal risk and skin cancer risk, Srivastava et al. (2019) matched people's fairness choices with different notions of group fairness: Demographic Parity, Error Parity, False Discovery Rate Parity, False Negative Rate Parity, False Positive Parity, and False Omission Parity. Their results showed that demographic parity best matched the fairness choices made by the majority of respondents in both scenarios. Thus, when it comes to group fairness, people favored algorithms that aim to equalize the positive rate across different groups. For instance, if 10 percent of all applicants to a university get admitted, this rate should be equal for all gender groups. The authors further found that in high-stakes situations, respondents weigh accuracy higher and inequality lower (Srivastava et al., 2019), although a qualitative study by Koene et al. (2017)

contradicted this finding, revealing that participants deemed ethical considerations more important than higher accuracy.

Two other studies compared perceptions of different formal definitions of fairness in the context of loan decisions (Kasinidou et al., 2021; Saxena et al., 2020). Both studies used the same scenarios to test three different fairness models: equal distribution ('money is split equally among candidates'), meritocratic distribution ('all the money is distributed to the candidate with the highest payback rate'), and calibrated/proportional distribution (money is split proportionally to candidates' payback rates'). The results indicated that people perceived the calibrated model to be the fairest.

Using predictive policing as a case study, Grgić-Hlača, Zafar et al. (2018) focused on different formal definitions of procedural fairness conceived as the selection of input features. The authors distinguished among three definitions of process fairness: feature-a priori fairness (a feature is perceived as fair, independent of its effect on the outcome); feature-accuracy fairness (a feature is perceived as fair if it increases the accuracy of an algorithm); and feature-disparity fairness (a feature is perceived as fair even if it increases disparity in the outcomes of an algorithm). The results indicated that respondents perceived feature-accuracy fairness to be the fairest process, followed by feature-apriori fairness and then feature-disparity fairness.

Furthermore, a study by Koene et al. (2017) investigated five different fairness concepts using the allocation of coursework topics among students. The results showed that the respondents could not agree on an algorithm, and the authors concluded that "the different fairness criteria users adopt can conflict with each other making it impossible for an algorithm to guarantee them simultaneously" (Koene et al., 2017, p. 2). Another study shed light on the trade-offs between different incompatible fairness definitions. Using pairwise comparisons, Harrison et al. (2020) investigated people's perceived fairness of algorithms in the criminal justice system that equalize false positive rates, accuracy, and outcome. The results indicated that respondents favored an algorithm that equalizes the false positive rate between groups over one that equalizes accuracy.

However, the technical design of an algorithm refers not only to the decision outcome but also to the decision process. Six studies investigated the perceived fairness of input features. Grgić-Hlača, Redmiles et al. (2018) identified eight feature properties that determine whether or not people perceive the use of said feature in an ADM system to be fair: reliability, relevance, privacy, volitionality, causes outcome, causes vicious cycle, causes disparity in outcomes, and caused by sensitive group membership. The results indicated that *relevance*, *causes outcome,* and *reliability* are most important for respondents when deciding whether it is fair to use a feature in an ADM system.

As expected, other studies found that respondents perceived features that directly relate to the issue at hand to be fairest and perceived unrelated features to be the most unfair (Grgić-Hlača, Zafar, et al., 2018; Van Berkel et al., 2019). In line with this, Plane et al. (2017) found that discriminatory advertising (e.g., better job ads) was evaluated as more problematic when predictions were made based on demographics instead of online behavior. Furthermore, respondents deemed the use of race and political affiliation as input features to be more unfair than age and health status. Qualifying this finding, Nyarko et al. (2020) discovered that respondents were generally averse to including sensitive features, like race and gender, in an ADM system for pretrial risk assessment; however, after respondents were told that including these features in the model can lead to better outcomes for minority groups, support for such 'non-blind algorithms' increased substantially. In one of the few studies conducted in a non-

Western context, Sambasivan et al. (2021) further found that missing data and misrepresentation of subgroups in the existing data were critical reasons for algorithmic unfairness to be perceived.

Another set of studies looked at explanations for a decision made by an algorithm as a critical aspect of perceived procedural fairness. In a study on AI-based news recommender systems, explanations for a decision made by an ADM system significantly increased respondents' perceptions of fairness (Shin, 2021). Qualifying this finding, Lee, Jain et al. (2019) suggested that outcome explanations had a significant influence on people's perceptions of algorithmic fairness, but the direction of the effects largely depended on the context. When explanations helped respondents understand uneven distributions, perceived fairness decreased. However, when explanations helped respondents understand equality in utility distribution, perceived fairness increased.

In terms of different explanation styles, the literature offers very nuanced results. Binns et al. (2018) found that when people were exposed to multiple explanation styles, *case-based* explanations (presenting similar cases to the cases being explained) had a negative influence on perceived fairness, especially compared to *sensitivity-based* explanations (showing how values of input features would have to change for a different classification). However, when exposed to only one explanation style, they found no substantial differences between the groups. Employing a similar methodology, Dodge et al. (2019) concluded that sensitivity- and case-based explanations more effectively exposed fairness discrepancies between different cases, while *input influence* (presenting all input features and their impact in the classification) and *demographic* explanations (offering information about the classification for individuals in the same demographic categories) increased respondents' confidence in understanding the model, leading to higher perceived fairness.

Looking at transparency more generally, Vallejos et al. (2017) found that young people demanded more information about an algorithm to perceive it as fair. However, empirical, experimental evidence offers ambiguous results. While Wang (2018) found that algorithmic transparency increased perceptions of fairness, Wang et al. (2020) observed that different degrees of transparency had no significant effect on algorithmic fairness.

**5.5** *Human predictors of perceived algorithmic fairness*

Another strand of empirical studies investigates human predictors of perceived algorithmic fairness. Only a few studies found an impact of sociodemographic variables. For instance, two studies suggested a significant influence of gender (Grgić-Hlača et al., 2020; Pierson, 2017); female respondents opposed gender as an input feature more strongly than male respondents did. While Helberger et al. (2020) did not find a significant effect of gender, their results indicated that education and age affected both perceptions of algorithmic fairness and people's reasons for such perceptions. In terms of age, young people perceived the inclusion of a sensitive feature in ADM systems to be unfair (Grgić-Hlača et al., 2020) and seemed to demand global approaches to regulating fairness measures and ethical guidelines for algorithms (Vallejos et al., 2017). Furthermore, Wang et al. (2020) found an interaction effect of education and self-interest: respondents with lower education levels perceived the outcome of an algorithmic decision as fairer when they benefited from it, compared to respondents with higher education levels, whose fairness perceptions were stable regardless of their personal benefit.

Moreover, two studies found that perceived algorithmic fairness hinged on self-interest, indicating that people tend to perceive algorithms as fairer when the ADM yields a positive outcome for them (Grgić-Hlača et al., 2020; R. Wang et al., 2020). In addition, Lee, Jain et al.

(2019) presented evidence that the difference between respondents' predicted outcomes versus the actual outcomes is associated with perceived algorithmic fairness: the greater the gap between prediction and actual outcome, the more unfair an algorithm was perceived.

Another aspect that received considerable attention is related to people's familiarity with data and algorithms. Saha et al. (2019) found that respondents who understood the mathematical definition of the fairness concept were more likely to reject it. However, evidence on the role of knowledge about AI is mixed. While Lee and Baykal's (2017) results indicated that higher levels of computer programming knowledge were associated with lower levels of perceived algorithmic fairness, Araujo et al. (2020) found that the effects ran in the opposite direction. Additionally, people's online self-efficacy—i.e., their belief that they are able to protect their data—tended to lead to more perceived fairness (Araujo et al., 2020). In relation to this, the same study presented empirical evidence that general concerns about how personal data is collected and used have a negative effect on respondents' perception of algorithmic fairness (Araujo et al., 2020).

Three other studies looked at political ideology as a predictor of algorithmic fairness perceptions. In two studies, Grgić-Hlača and colleagues (Grgić-Hlača et al., 2020; Grgić-Hlača, Redmiles, et al., 2018) found that conservative users perceived the inclusion of sensitive features such as gender and race in an ADM system as fairer than liberal users did. More specifically, if a feature increased disparity, liberal users' perceived fairness decreased substantially more than conservatives' perceived fairness (Grgić-Hlača, Redmiles, et al., 2018). Woodruff et al. (2018) also found that respondents saw a connection between unfairness in algorithms and national dialogues on ethnic and economic inequality. Based on these results, the authors concluded that fairness should be integrated as a value in the design and development of an algorithm.

## 5.6  Comparative effects (HDM vs. ADM)

In the literature on algorithmic appreciation and algorithmic aversion, many empirical studies investigated how individuals react to human compared to algorithmic decision-makers (Dietvorst et al., 2015; Logg et al., 2019). However, those studies did not investigate fairness perceptions specifically. Our literature review found fourteen studies that explicitly examined whether decisions made by humans or those by algorithms are perceived to be fairer, with ambiguous results. In one study, after being asked, "Who would, according to you, make a fairer decision: a human or artificial intelligence/computer?", 54% of respondents answered that they believed AI makes fairer decisions (compared to 33% for humans) (Helberger et al., 2020). This general finding was replicated by Marcinkowski et al. (2020); looking at university admission decisions, they found that ADM is perceived as fairer than HDM both in terms of distributive fairness and procedural fairness.

However, several studies provided different evidence that suggests that decisions and decision processes made by humans are viewed as fairer compared to algorithms, in the criminal justice system (Harrison et al., 2020; A. J. Wang, 2018), work-related decisions, such as hiring (Acikgoz et al., 2020; Newman et al., 2020) and social division tasks (M. K. Lee & Baykal, 2017). The main reason for this finding is that respondents believed that algorithms do not take qualitative information or context into account. However, Wang (2018) found that respondents were more willing to accept discriminatory outcomes when the outcomes were attributed to algorithms rather than to humans. Another experimental study with two country samples (German & US) found evidence for both positions, as HDM and ADM had different effects on different proxies of procedural fairness (Kaibel et al., 2019): while algorithms were rated higher in terms of consistency, human agents were rated higher in terms of personableness. Then again, other studies found no significant differences in perceived fairness between HDM and ADM (Plane et al., 2017; Suen et al., 2019).

The inconsistency of the empirical evidence suggests that fairness perceptions of HDM vs. ADM are highly context-dependent. Consequently, several studies looked at conditional effects or distinguished between different kinds of decisions. In line with conditional effects, an experiment on automated vs. human decision-making in policing suggested that procedural fairness perceptions hinge on social group representation (Miller & Keiser, 2020). In this US sample, black respondents rated ADM as fairer—but only when they felt their social group lacked representation in the HDM condition.

In her seminal study comparing different types of decisions, Lee (2018) found that for tasks requiring mechanical skills, ADM and HDM were perceived as equally fair. However, for tasks requiring human skills, HDM was perceived as fairer than ADM, as algorithms are perceived to lack intuition and subjective judgment capabilities. Araujo et al. (2020) investigated decisions in the media, (public) health, and justice sectors and compared high- and low-impact decisions. Overall, the findings suggested no significant differences between ADM and HDM in perceived fairness across the media, health, and judicial contexts; however, ADM was perceived as fairer than HDM in high-impact decisions in the health and justice sectors. In contrast to this finding, Nagtegaal (2021) distinguished between high- and low-complexity decisions and found that HDM was perceived as fairer than ADM in high-complexity tasks, because algorithms did not account for particular circumstances. However, for low-complexity tasks, algorithms were viewed as fairer, because humans are subjective and biased while computers were perceived as more objective.

Another strand of literature went beyond the binary distinction between HDM vs. ADM and also included hybrid forms of decision-making. For instance, Nagtegaal (2021) found that decision-making systems involving algorithms and humans are perceived as the most fair in high-complexity situations. However, the results from Newman et al. (2020) suggested that while algorithmic decisions with human oversight increased fairness perceptions, it was still outranked by pure HDM. Providing more information on the involvement of humans in the ADM process, Lee et al. (2019) allowed respondents to adjust the algorithmic allocation of resources and thereby overrule decisions made by the ADM system. The results showed that including outcome control in the decision-making process increased people's perceptions of algorithmic fairness. In another study, Wang et al. (2020) indirectly investigated the effect of humans-in-the-loop by adjusting the development and deployment of an algorithm to have more and less human involvement. However, the results showed no significant differences across the experimental conditions.

### 5.7 *Consequences of perceived fairness*

Only a few studies investigated the implications of the perceived fairness of algorithms. Shin and Park (2019) found that perceived fairness has a significant positive impact on satisfaction with algorithms that even exceeds the effect of perceived transparency. The effect was further moderated by people's levels of trust in algorithms. Several studies focused on the relationship between perceived algorithmic fairness and trust (Kasinidou et al., 2021; Shin, 2020, 2021; Shin et al., 2020). While Kasinidou et al. (2021) found no correlation between the two variables, causal analyses by Shin and colleagues (2020, 2021; 2020) suggested that fairness perception had a positive effect on trust in a recommendation made by an algorithm. The authors concluded that perceptions of fairness play an essential role in the adoption of algorithms.

With regard to using AI in human resources decisions (such as hiring), empirical evidence suggests that perceptions of procedural and interactional algorithmic unfairness were associated with lower organizational attraction or commitment and lower job pursuit intention (Acikgoz et al., 2020; Newman et al., 2020). Furthermore, low levels of perceived interactional fairness increased the likelihood of pursuing litigation against a company using ADM systems (Acikgoz et al., 2020).

In a survey on ADM in university admissions, Marcinkowski et al. (2020) tested how fairness perceptions of algorithmic decision processes and decision outcomes influence students' intentions to protest, students' willingness to exit, and the institution's reputation. The results yield three main insights. First, perceptions of distributive fairness and procedural fairness negatively influenced students' intention to protest against an ADM system. Second, only perceptions of procedural fairness had a negative effect on students' likelihood of exiting a university that bases their admission systems on ADM. Third, only perceptions of distributive fairness had a positive effect on the university's reputation. The authors concluded that the implementation of ADM systems in university admissions needs to account for the detrimental implications that perceived unfairness might cause. Using a qualitative research design, Woodruff et al. (2018) asked respondents belonging to traditionally marginalized groups about possible consequences of perceived algorithmic unfairness. The results indicated that "algorithmic fairness (or lack thereof) could substantially affect their trust in a company or product" (p. 1).

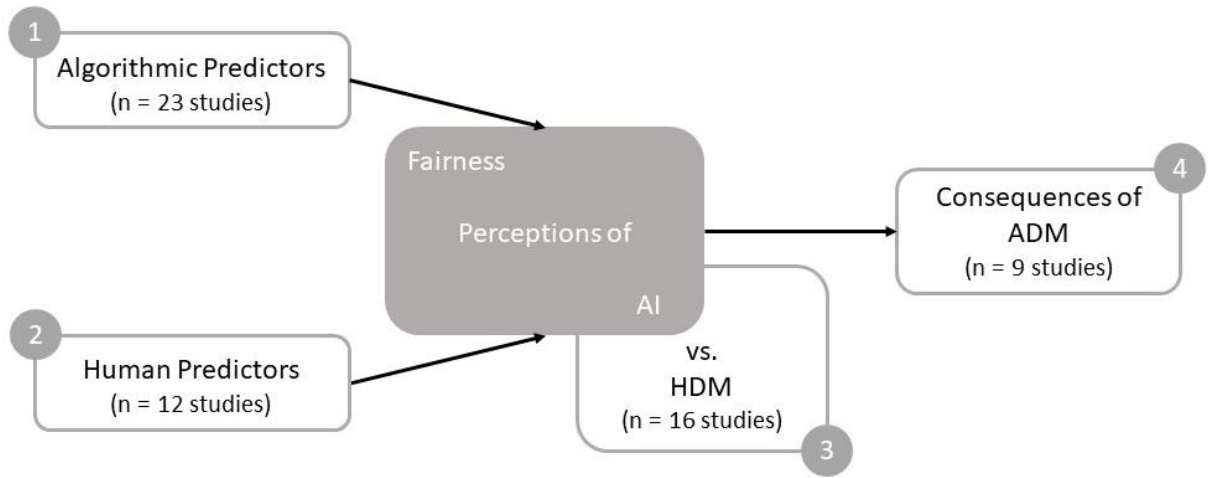

**Fig. 2.** Summary of the results in a process model

Four main insights can be summarized. First, the results of studies investigating algorithmic predictors of perceived algorithmic fairness indicate that preferences for different distribution norms are highly context-dependent and can vary substantially across domains, tasks, and algorithmic designs. However, respondents seemed to favor more straightforward fairness definitions over more complex ones. The literature also investigated how the process of ADM should be designed, yielding tentative evidence for a positive effect of explanations. Second, while studies on human predictors deliver inconclusive results in terms of sociodemographic variables, they also indicate that political ideology and self-interest influence citizens' fairness perceptions of ADM. Third, studies comparing fairness perceptions of HDM vs. ADM reveal that despite considerable variance in the investigated areas of application (e.g., justice, health, education system), the empirical results are ambiguous; fairness perceptions are highly context-sensitive, which makes generalizations about the perceived fairness of HDM vs. ADM infeasible. Fourth, while surprisingly little empirical research examines the consequences of perceived algorithmic (un)fairness, initial empirical insights reveal negative effects for institutions using an ADM system if that system is perceived to be unfair, especially with respect to its reputation and people's willingness to exit the institution.

## 6 Discussion

While the literature on algorithmic fairness has long been dominated by computer science studies in an attempt to mitigate bias and discrimination from machine learning models, there has been a recent push for social science research that takes a human-centric approach. This type of research puts the fairness perceptions of those most affected by ADM on the map. As Barabas et al. (2020, p. 174) stated: "As social scientists have long argued, the fairness of a data science project extends far beyond the technical properties of a given model. Data science is a socio-technical process, and the designers of algorithmic systems must embrace a process-driven approach to algorithmic justice." While earlier review papers have addressed human trust in AI (Glikson & Woolley, 2020), algorithmic discrimination in human resources (Köchling & Wehner, 2020), Big Data discrimination (Favaretto et al., 2019), and formal definitions of algorithmic fairness (Verma & Rubin, 2018), this systematic literature review is the first to shed light on perceptions of algorithmic fairness. It reveals the insights of the empirical literature on citizens' perceived fairness of ADM systems. In the following sections, we draw conclusions, identify blind spots, and outline promising avenues for future research.

### 6.1 *Theoretical Groundwork*

A key takeaway of this review is that the perceived fairness of ADM systems is highly context-dependent. Fairness perceptions are determined not only by the technical design of the algorithm but also by the area of application (e.g., pretrial risk assessment, hiring, university admission) and the specific task at hand (e.g., high-stakes vs. low-stakes). However, despite the variety of tasks examined in the literature, we argue that some of the inconclusiveness of the empirical results can also be attributed to the lack of coherent theoretical frameworks for perceived algorithmic fairness.

Two aspects stand out. First, as mentioned earlier, existing theoretical fairness approaches are not used consistently in the literature. For instance, while some authors define procedural fairness as input features (e.g., Grgić-Hlača, Zafar, et al., 2018), others conceive of it in broader terms that also include aspects of consistency and revocability (Marcinkowski et al., 2020). Second, most studies rely on fairness concepts that have been developed for decisions made by humans, such as in the context of organizations (Greenberg, 1990). However, empirical evidence suggests that people based their evaluations of ADM on different factors than the ones they use to evaluate HDM (Dietvorst et al., 2015). It is likely that people also include other criteria in evaluating the fairness of ADM and HDM. Following these arguments, we echo the call voiced by other authors (Lepri et al., 2018; Wong, 2020) for more theoretical groundwork on human perceptions of algorithmic fairness. Two avenues seem fruitful for future research: systematizing contextual factors that affect fairness perceptions of ADM; and unifying theoretical approaches concerning perceived algorithmic fairness.

### 6.2  *Diversification vs. Harmonization of Research*

We argue that a need exists for both diversification and harmonization of empirical research on algorithmic fairness perceptions. First, while the descriptive results show a large variety of research methods, they also reveal that the studies included in this review were almost exclusively conducted in Western democracies, predominantly the US. The problematic aspects of generalizing so-called *WEIRD* (White, Educated, Industrialized, Rich, Democratic) samples have been extensively discussed in psychological research (e.g., Henrich et al., 2010). The seminal work of Henrich et al. (2010) showed that fairness in decision-making is considerably dependent on the sociocultural context. It is likely that the same applies to fairness perceptions of algorithms. For instance, much cross-national variance exists globally in citizens' perceptions of AI (Kelley et al., 2019). Diversification of countries under investigation—especially those in which ADM systems are already widely implemented, such as China and South Korea—would greatly enrich the existing literature[2] (Sambasivan et al., 2021).

In addition, we also call for more diversification in terms of the investigated domains and tasks. Thus far, the literature has been dominated by fairness perceptions around ADM in the criminal justice system (especially the recidivism risk prediction system COMPAS) and ADM in human resources (especially hiring). However, the implementation of ADM systems has recently surged in many other areas of society, such as distributing social benefits (Noriega-Campero et al., 2020) and assigning corruption risk scores to public officials (Marzagão, 2017). This review reveals that fairness perceptions vary considerably across different algorithmic designs and different application areas (Araujo et al., 2020; M. K. Lee, 2018). Thus, more empirical research systematically comparing fairness perceptions across various domains and tasks is needed.

Second, given the plethora of different measurements used in the literature (see Table 3 in the supplementary materials), we argue in favor of more harmonization in this respect. Ideally, reliable measurements of perceived algorithmic fairness should be developed and validated following the theoretical groundwork outlined above. Not only would these make new findings more comparable, but they would also allow for more nuanced interpretations of the results. For instance, a multi-dimensional measurement for procedural fairness could adapt the six criteria (consistency, bias-suppression, accuracy, correctability, representativeness, ethicality) introduced by Leventhal (1980). Such a measurement would provide information about the specific deficiencies of an algorithmic process instead of merely indicating that a process is perceived as (un)fair. For example, knowing whether consistency concerns or accuracy concerns drive a perception of unfairness could enable researchers, developers, and policymakers to fine-tune ADM processes according to citizens' desires.

---

[2] Due to the selection of search strings and databases for this review, we cannot make any assertions about studies on perceived algorithmic fairness in languages other than English.

Recent studies have tackled this challenge for ADM-adjacent concepts, such as 'threats of AI' (Kieslich et al., 2021), introducing validated scales that are easily adaptable for use in different domains. Thus, new opportunities for interdisciplinary research collaborations are emerging. Survey experts from the social sciences can team up with computer scientists who have expertise in designing real (instead of hypothetical) ADM systems. Such collaborative studies would also answer the call prominently voiced by Rahwan et al. (2019) for more research on machine behavior.

### 6.3  *Understanding Consequences of ADM Systems*

Only few empirical studies investigated the consequences of algorithmic (un)fairness perceptions. This is surprising, in that actions or attitude changes resulting from perceived (un)fairness may have profound ramifications for institutions implementing ADM systems. Such systems are introduced for several reasons, including reducing costs, increasing impartiality, or improving decision outcomes. However, existing examples showed that ADM systems often come with unintended consequences. Citizens' perceptions of fairness may play a key role here.

Imagine the following scenario: A company uses ADM in its hiring process to select the most qualified applicants. A lack of knowledge about potential adverse effects might quickly backfire. For instance, if highly-qualified applicants view such a system to be unfair and consequently refrain from applying in the first place, the losses may be an obstacle to the desired goals. Conceptualizing perceptions of algorithmic fairness as a mediator variable in complex structural equation models has proven to be a useful way to study both the drivers and the consequences of algorithmic fairness (Shin, 2021; Shin & Park, 2019). We encourage more research that builds on such approaches and examines the potential implications of perceived algorithmic unfairness for other attitudinal and behavioral variables, especially in high-stakes situations.

### 6.4  *Understanding Emerging Trade-offs*

All algorithms face the challenge of trying to achieve many, sometimes incompatible goals simultaneously (M. S. A. Lee & Floridi, 2020). For instance, maximizing accuracy and fairness is often infeasible, as including sensitive features (such as race or gender) in the model tends to lead to more accurate results (Harrison et al., 2020). These are not the only algorithmic design considerations that can offset each other: adding transparency, accountability, explainability, privacy, and security concerns results in multifaceted matrices of trade-offs (Köbis et al., 2021).

Understanding how citizens evaluate the emerging trade-offs in algorithmic design in terms of fairness is highly relevant. How do citizens want those trade-offs to be solved by developers? In which configuration of trade-offs would citizens accept, trust, or support the use of an ADM system, and in which formats would they reject or actively oppose it? Again, as this review has shown, algorithmic fairness perceptions are highly context-dependent; thus, in future research, it is worth exploring whether and how the fairness of emerging trade-offs is evaluated in relation to different domains and tasks. For instance, people might favor accuracy in high-stakes medical algorithmic decisions but favor fairness for the distribution of social benefits. Understanding the emerging trade-off matrices should inspire future research that uses conjoint experiments, which would allow the consideration of multiple, incompatible goals jointly.

### 6.5 *From Human-in-the-Loop to Society-in-the-Loop*

Extending the literature on algorithmic appreciation (Logg et al., 2019), several studies included in this literature review investigated the perceived fairness of HDM vs. ADM. Here too, the results are inconclusive; they provide evidence for and against the assumption that algorithms are seen as more objective, impartial, and thus fairer than human decision-makers.

This variation of results can be found across different domains (e.g., recidivism risk prediction vs. hiring) and across different tasks within the same domain (e.g., two different hiring algorithms). This shows that fairness perceptions are highly context-specific and that every algorithm requires thorough investigation before being widely used. However, simply differentiating HDM and ADM falls short of capturing the complexity of the natural world (Binns, 2020). In most real-life tasks, ADM systems do not decide entirely on their own; instead, humans are also involved in the decision-making process at some point. For instance, COMPAS makes a prediction, which serves as a recommendation for the final decision by a human judge. Such forms of hybrid decision-making (Starke & Lünich, 2020) can be quite fine-tuned and deserve more nuanced attention from empirical literature.

We can conceive of three different approaches to human oversight to illustrate this point (Artificial Intelligence High-Level Expert Group, 2019). In the first, humans are involved in every step of the decision cycle of an ADM system and can intervene at any point (human-in-the-loop). In the second, humans can intervene during the design cycle of an ADM system and monitor the system's operation (human-on-the-loop). In the third, humans oversee the overall economic, societal, legal, and ethical impacts of an ADM system and have authority over its use in any situation (human-in-command). Fairness perceptions are likely to vary considerably among these cases of hybrid decision-making.

This leads to a broader point of institutional implementation. Whether it is in the criminal justice system, in companies, or in the education sector, introducing an ADM system into an institutional context raises a plethora of critical questions that extend far beyond the algorithmic design (e.g., input data, code) and the specifics of human involvement in the decision-making process (Wong, 2020). In this area, this review revealed a blind spot in the existing literature, as few studies considered the broader institutional context and the emerging questions in implementing an ADM system. For instance, should an ADM system replace a current decision system only because it is technologically possible? How are humans who might use ADM systems within an institution trained? How are decisions made by an algorithm communicated to those most affected by the decision (e.g., job applicants, defendants)—are they even made aware that the final decision was made by an algorithm? Do affected citizens have an opportunity to appeal, and who should ultimately be liable for false and/or discriminating classifications?

These are just a few examples of the multitude of emerging questions that are likely to profoundly affect citizens' fairness perceptions. Addressing these blind spots requires more interdisciplinary research involving computer scientists, social scientists, legal scholars, and ethicists (Lepri et al., 2018). Research on algorithmic fairness should also include the entirety of society, incorporating civil society organizations, high-level decision-makers in institutions, and public opinion in general. Thus, we argue that literature on the perceived fairness of ADM systems should adopt the society-in-the-loop framework, which Rahwan (2018) defined as human-in-the-loop plus social contract.

This approach allows for a more nuanced understanding of the complex intricacies of using algorithms for critical social decisions, and this framework also has practical implications. It addresses the limits of a narrowly-defined human-centric approach, arguing that ADM systems should be designed based on human preferences. Several studies in this review indicated that citizens tend to view relatively simple formal fairness definitions, such as demographic parity, as the most fair (Srivastava et al., 2019), arguably because other fairness definitions (such as error parity) are too complex for a lay audience. Indeed, telling respondents that excluding sensitive features might lead to detrimental outcomes for minority groups increases their support for more complex algorithms (Nyarko et al., 2020). Thus, designing ADM systems solely according to popular opinion might backfire. Instead, multiple stakeholders should be involved in the processes of designing such systems, establishing common ethical standards for algorithms, and integrating those standards into institutional and legal frameworks (Lepri et al., 2018; Žliobaitė, 2017).

# 7   Conclusion

As algorithmic decision-making increasingly penetrates all sectors of society, concerns about the fairness of such systems have arisen. As scholars and policymakers have demanded a human-centric approach to designing and implementing ADM, the empirical literature on perceived algorithmic fairness is surging. This systematic literature review crystallizes the insights of 39 empirical studies along four dimensions: algorithmic predictors, human predictors, comparative effects (HDM vs. ADM), and consequences of ADM. In conclusion, we call for more research from non-Western contexts, along with more theoretical and methodological groundwork to harmonize concepts and measurements of algorithmic fairness perceptions. Finally, we advocate for more interdisciplinary research adopting a society-in-the-loop framework, as we hope this work will contribute to fairer and more responsible algorithmic decision-making.

*Table 3: Overview of included articles (1/5)*

| Author, Year, Country of Data Collection | Method | Use Case | Fairness Concept | Fairness Measurement | N | Perceived fairness of algorithmic Design | Individual effects on perceived fairness | Subsequent effect of perceived fairness | Fairness perceptions of ADM vs. HDM |
|---|---|---|---|---|---|---|---|---|---|
| Acikgoz et al. 2020, USA | Quantitative Study | Hiring decisions | Procedural fairness, interactional fairness | Short multi-item scale (5-point Likert scale) based on Bauer et al., 2001 | N = 545 | – | – | ✓ | ✓ |
| Araujo et al. 2019, Netherlands | Quantitative Study | Multiple use cases (high & low stakes), news recommendations, Health, Justice | NA | Single-item measure (7-point Likert scale) | N = 958 | – | ✓ | – | ✓ |
| Binns et al. 2018, UK | Mixed Methods | Loan decisions, Promotion at work, Car insurance pricing, transportation, freezing of finances | Distributive, procedural, interactional, informational fairness, | Short multi-item scale (5-point Likert scale) based on Colquitt & Rodell, 2015 | N = 409 | ✓ | ✓ | – | – |
| Dodge et al. 2019, USA | Quantitative Study | COMPAS | NA | Short single-item scale (7-point Likert scale) | N = 160 | ✓ | – | – | – |
| Grgić-Hlača, Redmiles et al. 2018, USA | Quantitative Study | COMPAS | Distributive & procedural fairness | Single-item measure (7-point Likert scale) | N = 576 | ✓ | ✓ | – | – |
| Grgić-Hlača, Zafar et al. 2018, USA | Quantitative Study | Predictive policing & COMPAS | Distributive & procedural fairness | Short multi-item scale | N = 200 | ✓ | – | – | – |
| Grgić-Hlača, Weller, & Redmiles, 2020, USA | Quantitative Study | COMPAS | Procedural fairness | Short multi-item scale (7-point Likert scale) based on Grgić-Hlača et al., 2018a | N = 203 | – | ✓ | – | – |
| Harrison et al. 2020, USA | Mixed Methods | COMPAS | Group fairness, procedural fairness | Single-item measure (5-point Likert scale) | N = 502 | ✓ | – | – | ✓ |
| Helberger et al. 2020, Netherlands | Mixed Methods | NA | Procedural fairness, distributive fairness, substantive fairness | Content analysis | No survey | – | ✓ | – | ✓ |

*Table 3: Overview of included articles (2/5)*

| Author, Year, Country of Data Collection | Method | Use Case | Fairness Concept | Fairness Measurement | N | Perceived fairness of algorithmic Design | Individual effects on perceived fairness | Subsequent effect of perceived fairness | Fairness perceptions of ADM vs. HDM |
|---|---|---|---|---|---|---|---|---|---|
| Kaibel et al. 2019, Germany / USA | Quantitative Study | Hiring decisions | Procedural justice | Multiple sub-dimensions measured on different scales based on Bauer et al., 2001 | N = 183 | – | – | – | ✓ |
| Kasinidou et al. 2021, multi-national | Mixed Methods | Car insurance pricing, transportation, loan decision | Distributive fairness: equal distribution, meritocratic distribution, equity distribution | Single-item measure (5-point Likert scale) | N = 99 | ✓ | – | ✓ | – |
| Koene et al. 2017, NA | Mixed Methods | Coursework allocation | Multiple alternative concepts discussed | NA | NA | ✓ | – | – | – |
| Lee & Baykal 2017, USA | Mixed Methods | Resource allocation | Economic fairness, numerical equality, proportional equality, distributive justice, outcome fairness | Short multi-item scale (7-point Likert scale) | N = 166 | ✓ | ✓ | – | ✓ |
| Lee, Kim & Lizarondo 2017, USA | Qualitative study | Resource allocation | Equality, equity, distributive fairness | Explorative interview | N = 31 | ✓ | – | – | – |
| Lee 2018, USA | Quantitative study | Managerial decisions | Procedural fairness | Single-item measure (7-point Likert scale) | N = 321 | – | – | – | ✓ |
| Lee, Kusbit et al. 2019, NA | Qualitative study | Resource allocation | Equality, equity, distributive fairness | Explorative multi-stage study | N = 24 | ✓ | – | – | – |
| Lee, Jain et al. 2019, USA | Mixed methods | Division of goods | Informational fairness, distributive & procedural fairness | Short multi-item scale (7-point Likert scale) | N = 71 | ✓ | ✓ | – | ✓ |
| Marcinkowski et al. 2020, Germany | Quantitative study | University admission | Distributive & procedural Fairness | Single-item measure (5-point Likert scale) | N = 304 | – | – | ✓ | ✓ |
| Miller & Keiser 2020, USA | Quantitative study | Use of ADM for government decision-making | Procedural fairness | Single-item measure (5-point Likert scale) based on Sunshine & Tyler, 2003 | N = 394 | – | – | – | ✓ |

*Table 3: Overview of included articles (3/5)*

| Author, Year, Country of Data Collection | Method | Use Case | Fairness Concept | Fairness Measurement | N | Perceived fairness of algorithmic Design | Individual effects on perceived fairness | Subsequent effect of perceived fairness | Fairness perceptions of ADM vs. HDM |
|---|---|---|---|---|---|---|---|---|---|
| Nagtegaal 2021, Netherlands / UK | Mixed methods | Decision in Public Management | Procedural Justice | Short multi-item scale (7-point Likert scale) based on Lind & Tyler, 1988 | N = 235 | – | – | – | ✓ |
| Newman et al. 2020, USA | Quantitative study | human resource decisions | Procedural fairness | 4-item scale (7-point Likert scale) based on Conlon et al., 2004 | N = 2452 | – | – | ✓ | ✓ |
| Nyarko et al. 2020, USA | Quantitative study | Pretrial risk assessment | NA | Fairness not measured, but preferences on different algorithmic decision-making processes | N = 1009 | ✓ | – | – | – |
| Pierson 2017, NA | Quantitative study | Multiple use cases (COMPAS, recommendation systems, etc.) | NA | Single-item measure (7-point Likert scale) | N = 823 | – | ✓ | – | – |
| Plane et al. 2017, USA | Quantitative study | Targeted advertising | NA | Fairness not measured, but perceived problems with targeted advertisement within multiple sub-categories | N = 988 | ✓ | – | – | ✓ |
| Saha et al. 2019, USA | Quantitative study | Award / job distribution | Demographic parity | Comprehension, not fairness measured | N = 496 | – | ✓ | – | – |
| Sambasivan et al. 2021, multi-national (mainly India) | Qualitative study | NA | Cross cultural justice (e.g. restorative justice) | NA | N = 36 | ✓ | – | – | – |
| Saxena et al. 2020, USA | Quantitative study | Loan decisions | Distributive fairness: equal distribution, meritocratic distribution, equity distribution | Short multi-item scale (9-point Likert scale) | N = 2000 | ✓ | – | – | – |

*Table 3: Overview of included articles (4/5)*

| Author, Year, Country of Data Collection | Method | Use Case | Fairness Concept | Fairness Measurement | N | Perceived fairness of algorithmic Design | Individual effects on perceived fairness | Subsequent effect of perceived fairness | Fairness perceptions of ADM vs. HDM |
|---|---|---|---|---|---|---|---|---|---|
| Shin & Park 2019, NA | Mixed methods | Recommendation algorithms | NA | NA | N = 230 | ✓ | – | ✓ | – |
| Shin 2020, USA | Quantitative study | Recommendation of Digital Platforms | NA | Short multi-item scale (7-point Likert scale) based on Shin et al., 2020 | N = 391 | – | – | ✓ | – |
| Shin et al. 2020, South Korea | Quantitative study | Recommender systems | NA | Short multi-item scale (7-point Likert scale) | N = 345 | – | – | ✓ | – |
| Shin 2021, USA | Quantitative study | News recommendation | NA | Multiple sub-dimensions measured on different scales | N = 350 | ✓ | – | ✓ | – |
| Smith et al. 2020, NA | Qualitative interview | NA | NA | NA | NA | ✓ | – | – | – |
| Srivastava et al. 2019, USA | Quantitative study | Multiple use cases for risk prediction | Mathematical formulations of fairness, demographic parity, disparate impact, equality of odds, calibration | Approval of fairness models, not fairness itself measured | N = 220 | ✓ | – | – | – |
| Suen et al. 2019, China | Mixed methods | Hiring decisions | Procedural fairness | Short multi-item scale (5-point Likert scale) based on Guchait et al., 2014 | N = 180 | – | – | – | ✓ |
| Vallejos et al. 2017, UK | Qualitative study | NA | NA | Content analysis | N = 26 | ✓ | ✓ | – | – |
| van Berkel et al. 2019, USA | Qualitative study | COMPAS | Group fairness, individual fairness, equality of opportunity | Not fairness, but predictors (perceived as fair) measured | N = 90 | ✓ | – | – | – |
| Wang 2018, USA | Quantitative study | risk assessment in criminal justice | Procedural justice | Single-item measure (5-point Likert scale) | N = 10226 | ✓ | – | – | ✓ |

*Table 3: Overview of included articles (5/5)*

| Author, Year, Country of Data Collection | Method | Use Case | Fairness Concept | Fairness Measurement | N | Perceived fairness of algorithmic Design | Individual effects on perceived fairness | Subsequent effect of perceived fairness | Fairness perceptions of ADM vs. HDM |
|---|---|---|---|---|---|---|---|---|---|
| Wang, Harper & Zhu 2020, USA | Quantitative study | Master qualification rating on MTurk | Distributive & procedural fairness | Short multi-item scale (7-point Likert scale) based on Franke et al., 2014; Lee & Baykal, 2017 | N = 590 | ✓ | ✓ | – | ✓ |
| Woodruff et al. 2018, USA | Qualitative study | Multiple use cases | Individual perceptions | NA | N = 44 | – | ✓ | ✓ | – |

**Table 4.** Overview of databases.

| database | scope[3] | fields of research |
|---|---|---|
| Web of Science | about 12,000 journals > 160,000 conference proceedings | Science, Social Science, Arts & Humanities |
| PsycINFO | About 2,500 journals | Psychology and Social Science |
| IEEE Xplore | > 190 journals | Computer Science, Electrical Engineering and Electronics |
| Scopus | > 23,000 journals | Science, Technology, Medicine, Social Science, Arts & Humanities |

[3] The information about the scope and the fields of research was taken from the homepages of the electronic databases.